\documentclass{ab}  
\usepackage{graphicx,ulem}
\usepackage{txfonts}
\usepackage{color}
\begin{document}
  \titlerunning{ SS 433 eclipses}
\authorrunning{ Bowler}
   \title{SS 433: Stationary lines and primary eclipses }

   \subtitle{}

   \author{M. G.\ Bowler \inst{}}

   \offprints{\\   \email{michael.bowler@physics.ox.ac.uk}}
   \institute{University of Oxford, Department of Physics, Keble Road,
              Oxford, OX1 3RH, UK}

 
  \abstract
   {It is important to understand in as great detail as possible the properties of the Galactic microquasar SS\,433, which is likely the one and only local ULX. Knowledge of the the orbital velocity of the compact object depends largely on studies of a broad He II line;   the Balmer H$\alpha$ line and other stationary lines have a two horned structure, interpreted as emission from a circumbinary disk orbiting a system of total mass approximately 40 $M_\odot$. } 
  {To report certain results in which the variation of stationary spectra at the time of primary eclipses is important for better understanding of the properties of this unique system. }
   {Stationary H$\alpha$ and He I spectra, taken almost
     nightly over two orbital periods of the binary system, were analysed some years ago. Those spectra exhibit a two horned structure with intensity variations in excellent agreement with radiation from a circumbinary disk. There are, at first sight, some discrepancies at the time of primary eclipses. These discrepancies are examined and are relevant to primary eclipse data confirming that a particular doublet of C II emission lines is unambiguously associated with the compact object photosphere. From these data the speed of the compact object in its orbit has been extracted. }  
  { The principal discrepancy between the published stationary spectra and emission from a circumbinary disk is a trivial result of the normalisation adopted. In eclipse the continuum is halved and for about two days at eclipse the normalised spectra appear twice as intense as at other times. After this is taken into account, there remains only an imbalance in intensity from the approaching and receding sides of the circumbinary disk during a single eclipse. During eclipse the continuum is halved in intensity and so is C II intensity; in the two primary eclipses C II is eclipsed. The orbital speed extracted from these C II data is 176 km s$^{-1}$, in good agreement with the interpretation of the He II data. }


   \maketitle
%

\section{Introduction}

The Galactic microquasar  SS 433 is very luminous and unique in its continual ejection
of plasma in two opposite jets at approximately one quarter the speed
of light. The system is a 13 day binary, powered by supercritcal accretion by the compact member from its Companion. The orbital speed of the compact object seems fairly well established but, in order to determine the mass of the compact object, in addition either a measurement of the orbital velocity of the companion is needed or a measure of the total mass of the system. The real question is whether the compact object is a black hole and if so, a low or high mass stellar black hole. Since SS 433 may well be the one and only UltraLuminous X-ray source in the Galaxy, this question has gained renewed importance from optical spectra of extra-Galactic ULXs which look very like the optical emission spectra of SS 433 (Fabrika et al 2015).

I have written this note to place on record some inferences from changes in optical spectra of SS 433 at the time of primary eclipse. The first relates to the mass of the system inferred from interpretation of stationary lines in terms of radiation from a circumbinary disk; I discuss certain apparent discrepancies visible in the measured spectra over about two days of primary eclipse. The second inference is drawn from the behaviour of a C II emission doublet, which exhibits an oscillation of about 175 km s$^{-1}$ in phase with the compact object and is eclipsed for about two days, thus demonstrating its origin within the photosphere of the compact object.

\section{The circumbinary disk at eclipse}

\subsection{ The circumbinary disk}

   Stationary emission lines in the spectra of SS 433 display a persistent two horn structure of just the kind expected for emission from an orbiting ring, or disk, seen more or less edge on. The horn separation corresponds to a rotation speed in excess of 200 km s$^{-1}$ and attributed to material orbiting the centre of mass of the binary system implies a system mass in excess of 40 $M_\odot$. The H$\alpha$ spectra between JD 2453000 + 245 and +274 were originally discussed in  Blundell, Bowler \& Schmidtobreick (2008). Fig.2 of Schmidtobreick \& Blundell (2006) presents the stacked profiles of H$\alpha$ and the He I lines at 6678 and 7065 \AA\ . The interpretation of these profiles in terms of stimulation of a circumbinary disk by radiation from the compact object has been discussed in tedious detail by Bowler (2010, 2011a, 2013). The refined model of Bowler (2013) is extremely successful in accounting for the rather different profiles of H$\alpha$ and the He I lines. I here draw attention to the crucial properties, clearest in the He I emission because He I is most sensitive to proximity of the compact object to the circumbinary ring (Bowler 2013).
       The profiles sway from blue to red and back with a period of 13 days. Extreme blue occurs at approximately +244, +257 and + 270, when the compact object is in quadrature and moving towards us, Extreme red occurs at +252 and + 265, quadrature moving away. At days + 248, + 261 and + 274 the red and blue horns are of equal height and the lines highly symmetric. On these days the compact object passes directly in front of the companion and so this symmetry is to be expected. Such symmetry is also to be expected at primary eclipse and indeed is seen on days + 256 and + 269. 
       
 \subsection{Discrepancies real and apparent}
       
       These are actually about one day after primary eclipse. Days +254 and +255 span the first primary eclipse in these data; on these two days the blue horn is decidedly suppressed relative to the red. This is the only real discrepancy between the model of Bowler (2013) and these spectra. There is a slight discrepancy on days +267 and +268 but I do not regard this as significant.  With only two examples of primary eclipse in this data set, it might just be fluctuations in emission from the ring. 
       
       There is another discrepancy that is highly visible, but only apparent. The normalised intensities on days + 254, +255 and on + 267, 268, spanning the first and second primary eclipses, are about twice the typical intensities. This observation demonstrates that the intensity discrepancy is only apparent and also serves to confirm the ephemeris to well within a day. During primary eclipse the continuum background is reduced by about a factor of two; a convenient reference is Fabrika (2004). Because photometry was not available when the observations were made, the spectra in Schmidtobreick \& Blundell (2006) were normalised to a smoothed continuum. Thus the sources of H$\alpha$ and He I emission lines are not eclipsed, the ephemeris is confirmed and the apparent flaring of these lines at primary eclipse is illusory. To put it informally, these highly visible apparent flares are informative and no cause for alarm.

\section{On orbital speed of the compact object}
\subsection{What is known}

All attempts to calculate the mass of the compact object in SS 433 have relied upon the orbital speed being 175 km s$^{-1}$. If it were rather - say - 100 km s$^{-1}$ the mass estimates would be drastically changed. Thus in a second paper on INTEGRAL results (Cherepashchuk et al 2013) the authors state their opinion that "It is also important to improve the radial velocity curve from observations of emission lines (He II 4686, CII 7231, 7236, etc) in order to reliably determine the mass function of the compact star."

     The principal discussion of extraction of the orbital speed of the compact object from He II 4686 \AA\ is Fabrika \& Bychkova (1990). This He II line has a complex structure with both a broad and narrow component. The broad component is somewhat asymmetric and its Doppler shift agrees with that expected for a source in the compact object. It is almost totally eclipsed at the photometric primary eclipse. The profile of the He II line and its variation during the eclipse is discussed in detail in Goranskii et al (1997). The result given in Fabrika (2004) is 176 $\pm$13 km s$^{-1}$. If there are problems, they are likely to be in the complexity and width of the line, in particular the need to get rid of the influence of the relatively narrow component.
     
          So far as I know the first and (apart from the present note) only result from the C II doublet at 7231, 7236 \AA\ was obtained by Gies et al (2002). This doublet is sharp but it is weak and somewhat confused by terrestrial atmospheric absorption. Each run contains no more than 10 consecutive days, but the data are sufficient for Gies et al (2002) to have established that the (sinusoidal) velocity curve from the C II doublet is maximally red shifted when the compact object is receding at its fastest from us, given the ephemeris. The orbital speed was found to be 162 $\pm$ 29 km s$^{-1}$, consistent with the He II results.
          
             There is one other piece of evidence that indicates that the speed is greater than ~$\sim$130 km s$^{-1}$and probably about 175 km s$^{-1}$. It is a trifle unusual. The H$\alpha$ lines (and the He I lines) contain a broad component that is indubitably produced in the wind from the accretion disk (Blundell, Bowler \& Schmidtobreick 2008). The centroid oscillates approximately sinusoidally with an amplitude of $\sim$130 km s$^{-1}$. It is delayed by about one quarter of a period relative to the ephemeris (and the He II and C II doublet). This shift of phase is explained if the wind line is averaged over a few days of emission and the orbital velocity is approximately 175 km s$^{-1}$ (Bowler 2011b).
             
\subsection{Further results from C II data}

The Schmidtobreick/Blundell data set covers the spectral range for the stationary C II doublet and despite the weakness of the line and the data having been taken for an entirely different purpose, the data are useable. The averaged spectra can be found in Fig.1 of Schmidtobreick \& Blundell (2006). The daily spectra in this region have not been published but I have access to them. These data have a considerable advantage over those of Gies et al (2002) because they cover days +245 to +274, over two complete orbits, night by night. In particular they include two complete primary eclipses. Although the signal is weak and troubled by terrestrial lines, it is clear that the C II doublet is eclipsed over the two days of primary eclipse and the Doppler shifts of the doublet could be determined with an accuracy perhaps superior to that of Gies et al 2002. In these data the primary eclipses cover days 254.5, 255.5 and separately 267.5, 268.5. On just these days the neighbouring relatively feeble He I line at 7281 \AA\ doubles in intensity relative to the continuum background, just as the more prominent He I lines. The doublet from C II does not perceptibly change its intensity relative to the continuum. Thus their source is eclipsed to the same extent or greater than the continuum and is therefore associated with emission from the disk or its corona. It must share the orbital motion of the compact object. Fitting the Doppler shifts of the C II doublet over more than two complete periods I found an orbital speed of 176 $\pm$13 km s$^{-1}$.

\section{Conclusions}

The first conclusion is that the apparent outbursts visible in Fig.2 of Schmidtobreick \& Blundell (2006) are due to the normalisation of the spectra to the continuum that loses approximately half its intensity at primary eclipse. The data are not sufficient to investigate any residual discrepancies further. I conclude that here there are no grounds for suspecting the circumbinary disk description of these lines is substantially in error and hence that the inference of a system mass $\sim$ 40 $M_\odot$ has survived.
    My second conclusion is that the C II doublet is likely to give a true measure of the orbital velocity of the compact object. My value is 176 $\pm$13 km s$^{-1}$.

\begin{acknowledgements}
The spectra covering the C II doublet day by day are a small part of the remarkable Schmidtobreick/Blundell data set. They were made available to me over 7 years ago by K. M. Blundell. The observations were made in 2004 with the ESO NTT, with a grant of Director's discretionary time.

\end{acknowledgements}

\end{document}